\magnification=1095
\raggedbottom
\overfullrule=0pt
\headline={\ifnum\pageno>1 \hfil -- \folio\ -- \hfil \else\hfil\fi}
\footline={}
\font\smc cmcsc10 at 11 truept
 at 16 truept
 at 16 truept
\font\sml cmr10 at 10 truept
\font\stt cmtt10 at 10 truept

\def\largeskip{\vskip 0.5truein}

\def\hw{\hidewidth}
\def\ft#1{$^{\;\rm #1}$}

\def\Msol{\ifmmode{\rm M}_{\mathord\odot}\else M$_{\mathord\odot}$\fi}
\def\deg{\ifmmode^{\mathord\circ}\else $^{\mathord\circ}$\fi}

\def\ls{\lower 2pt \hbox{$\;\scriptscriptstyle \buildrel<\over\sim\;$}} 
\def\gs{\lower 2pt \hbox{$\;\scriptscriptstyle \buildrel>\over\sim\;$}}

\def\kms{km~s$^{-1}$}
\def\m#1{$^{-#1}$}
\def\asec{$^{\prime\prime}$}

\def\deg{$^{\circ}$}

\def\tten#1{$\times 10^{#1}$} 
\def\ten#1{$10^{#1}$}

\def\iii{~{\smc iii}}
\def\iv{~{\smc iv}}

\def\a{$\alpha$}
\def\b{$\beta$}

\def\s #1{$^{\;\rm #1}$}

\def\asca{{\it ASCA}}
\def\rosat{{\it ROSAT}}
\def\exosat{{\it EXOSAT}}
\def\einstein{{\it Einstein}}

\def\feka{Fe~K$\alpha$}
\def\s#1{SIS$\,#1$}
\def\g#1{GIS$\,#1$}

\newcount\Q
\Q=0

\input psfig
\newcount\F
\F=0
\advance\F by 1 \newcount\FHa     \FHa=\F     
\advance\F by 1 \newcount\Fspec   \Fspec=\F   
\advance\F by 1 \newcount\Fcont   \Fcont=\F   
\advance\F by 1 \newcount\Flimits \Flimits=\F 
\advance\F by 1 \newcount\FBLRGs  \FBLRGs=\F  

\newcount\T
\T=0
\advance\T by 1 \newcount\Tcounts \Tcounts=\T 
\advance\T by 1 \newcount\Tfit    \Tfit=\T    
\advance\T by 1 \newcount\TBLRGs  \TBLRGs=\T  

\font\bigsf=cmssbx10 scaled 1400
\hbox{
\psfig{figure=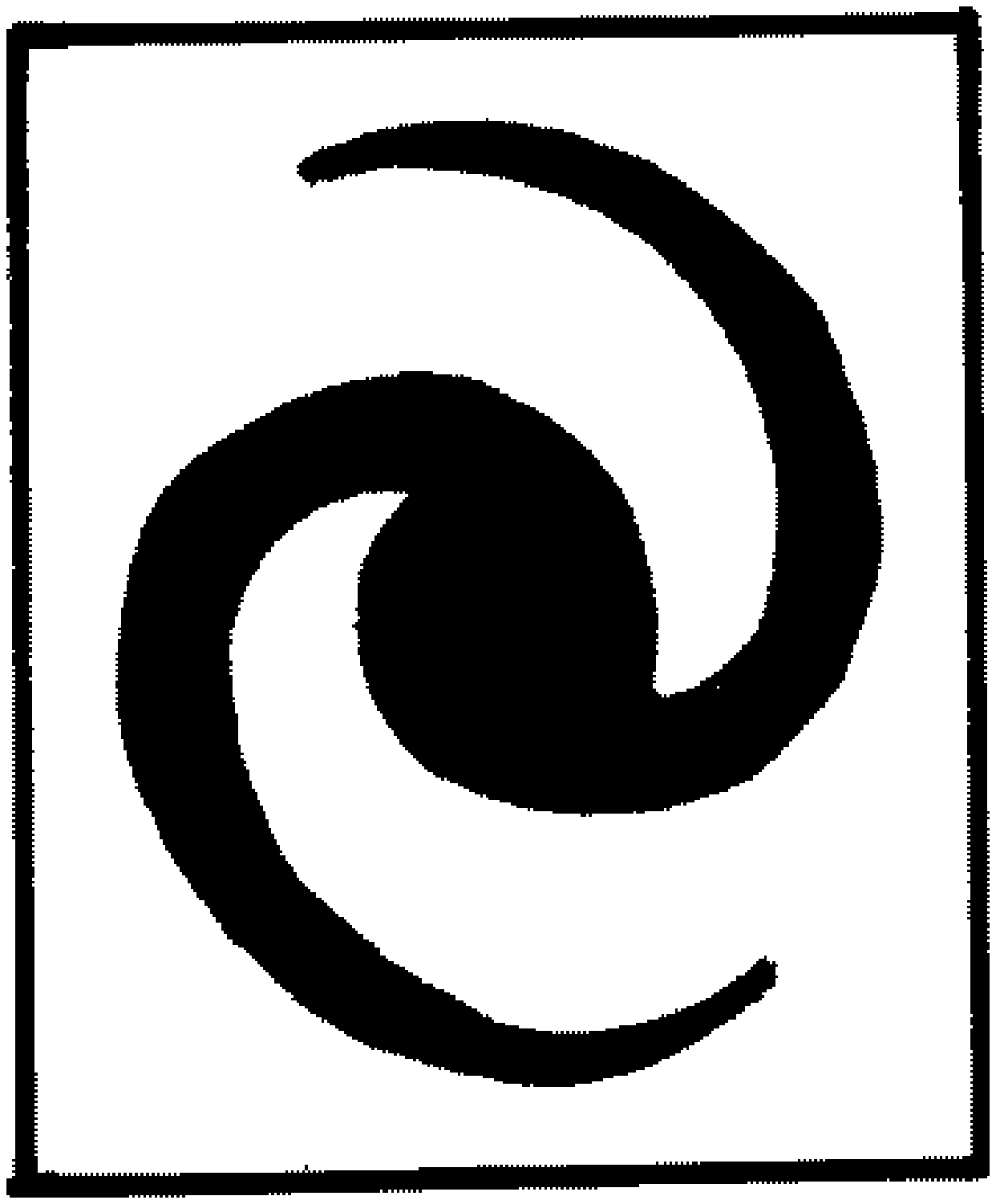,height=1in,width=1in}
\hskip 0.5truein 
\vbox to 1. truein {\bigsf \vfill
\noindent UNIVERSITY OF CALIFORNIA AT BERKELEY 
\medskip \noindent ASTRONOMY DEPARTMENT
\vfill}
}

\largeskip

\centerline{\bf THE {\it ASCA} X-RAY SPECTRUM OF THE BROAD-LINE RADIO GALAXY} 
\centerline{\bf PICTOR A: A SIMPLE POWER LAW WITH NO \feka\ LINE}

\largeskip
\centerline{\smc Michael Eracleous\ft{1,2} and Jules P. Halpern\ft{3}}

\footnote{}{\parindent=0pt \sml
\item{\ft{1}} Hubble Fellow
\item{\ft{2}} Department of Astronomy, University of California, Berkeley, 
CA 94720, {\stt mce@beast.berkeley.edu}
\item{\ft{3}} Columbia Astrophysics Laboratory, Columbia University, 550 
West 120th Street, New York, NY 10027, {\stt jules@astro.columbia.edu}
}

\largeskip
\centerline{To appear in {\it The Astrophysical Journal}}

\vfill

\centerline{ABSTRACT}
\bigskip
\centerline{\vbox{\hsize=6truein \noindent
We present the X-ray spectrum of the broad-line radio galaxy Pictor~A
as observed by \asca\ in 1996. The main objective of the observation
was to detect and study the profiles of the Fe~K$\alpha$ lines.  The
motivation was the fact that the Balmer lines of this object show
well-separated displaced peaks, suggesting an origin in an accretion
disk.  The 0.5--10~keV X-ray spectrum is described very well by a
model consisting of a power law of photon index 1.77 modified by
interstellar photoelectric absorption.  We find evidence for neither a
soft nor a hard (Compton reflection) excess.  More importantly, we do
not detect an Fe K$\alpha$ line, in marked contrast with the spectra
of typical Seyfert galaxies and other broad-line radio galaxies
observed by \asca. The 99\%-confidence upper limit on the equivalent
width of an unresolved line at a rest energy of 6.4~keV is 100~eV,
while for a broad line ({\it FWHM} $\approx 60,000~{\rm km~s^{-1}}$)
the corresponding upper limit is 135~eV.  We discuss several possible
explanations for the weakness of the \feka\ line in Pictor~A paying
attention to the currently available data on the properties of \feka\
lines in other broad-line radio galaxies observed by \asca.  We
speculate that the absence of a hard excess (Compton reflection) or an
\feka\ line is an indication of an accretion disk structure that is
different from that of typical Seyfert galaxies,
e.g., the inner disk may be an ion torus.
\bigskip\noindent
{\it Subject headings}: accretion, accretion disks -- galaxies: active --
galaxies: individual (Pictor~A) -- galaxies: nuclei -- line: profiles
-- X-rays: galaxies
}}

\vfill

\eject

\centerline{\smc 1. introduction}
\bigskip

The observed X-ray spectra of (radio-quiet) Seyfert galaxies have been
interpreted for almost a decade as a combination of a power-law
continuum which is seen directly and its reflection from cool, dense
matter (e.g., Pounds et al.  1987; Turner \& Pounds 1989; Nandra \&
Pounds 1994). The power-law continuum is generally thought to be the
primary radiation from an X-ray source associated with the inner parts
of the accretion disk while the reflecting medium is believed to be
the accretion disk itself -- the only structure in the vicinity of the
X-ray source thought to have a Thompson optical depth greater than
unity (George, Nandra, \& Fabian 1991; George \& Fabian 1991; Matt,
Perola, \& Piro 1991; but see also Nandra \& George 1994 for a
possible alternative, inspired by the ideas of Guilbert \& Rees).  The
reflected X-rays manifest themselves as a flattening of the spectrum
at energies higher than 10~keV, and as a number of fluorescent lines,
the most prominent of which is Fe~K\a\ at 6.4--6.9~keV (the exact
energy depending on the ionization state of the disk). The recent
detection of extremely broad ({\it FWZI} $\sim 0.3\,c$), asymmetric
\feka\ lines in the spectra of Seyfert galaxies by \asca\ bolsters the
above picture since their profiles conform to relativistic disk
kinematics (Mushotzky et al. 1995; Tanaka et al. 1995; Nandra et
al. 1997$a$). The observed X-ray spectroscopic properties have led to
models for the structure of the inner accretion disk involving a hot
corona overlaying an optically thick, geometrically thin disk (Haardt
\& Maraschi 1991, 1993; Haardt, Maraschi, \& Ghisellini 1994; Sincell
\& Krolik 1997). X-ray photons produced in the corona are reflected by
the underlying cool disk producing the reflected continuum in the
strength necessary to explain the observations.

The radio-loud analogs of Seyfert galaxies, the broad-line radio
galaxies (hereafter BLRGs) are considerably less numerous and hence
have not been studied as extensively. Nevertheless, several
indications suggest that the central engines of BLRGs (and possibly
their entire accretion flows) are systematically different from those
of Seyfert galaxies. The most obvious, and yet not duly appreciated,
difference is the ability of the central engines of BLRGs to
accelerate and collimate powerful relativistic jets. A more subtle,
but still significant, difference is that the {\it optical} emission
lines of BLRGs are about twice as broad as and considerably more
structured than those of Seyfert galaxies (Miley \& Miller 1979;
Steiner 1981). Unlike Seyfert galaxies, about 10\% of BLRGs feature
double-peaked Balmer lines which are characteristic of accretion disk
dynamics (Eracleous \& Halpern 1994). Moreover, the line-emitting gas
in the nuclei of BLRGs is thought to be preferentially confined to a
plane perpendicular to the axis of the radio jet (Wills \& Browne
1986; Jackson \& Browne 1990), and synthetic line profiles computed
under this assumption match the observed profiles quite well (Jackson,
Penston \& P\'erez 1991; Corbin 1997; Eracleous \& Halpern
1998$a$). From a theoretical perspective, it has long been suspected
that the inner accretion disks of BLRGs are hot, ion-supported tori\ft{3}
(Rees et al. 1982) rather than disk-corona sandwiches as proposed for
Seyferts. If so, only the outer disk at radii $r\gs 300\,r_{\rm g}$
($r_{\rm g}\equiv GM/c^2$) would be able to reflect the primary X-rays
with the main consequence that the reflected continuum and the Fe K\a\
lines should be considerably weaker in BLRGs than in Seyferts. Indeed,
the available data support this view. As shown first by Zdziarski et
al. (1995) and verified later by Wo\'zniac et al. (1998) the X-ray
spectra of BLRGs have considerably weaker and narrower \feka\ lines
and a reflected continuum consistent with zero (but see also our later
discussion in \S4.1). To reconcile the fact that \feka\ lines are
detected in the spectra of BLRGs but a reflected continuum is not,
Wo\'zniac et al. (1998) suggest that there is no medium in the
vicinity of the primary X-ray source that can act as an efficient
reflector. Instead, they propose that the \feka\ lines originate in
matter of moderate column density (of order \ten{23}~cm\m{2}) at a
large distance from the X-ray source, where reflection is inefficient.
\footnote{}{\parindent=0pt \item{\ft{3}} \sml The ion torus is very similar 
to what is known today as an advection-dominated accretion flow 
(Narayan \& Yi 1994, 1995)}

\topinsert
\centerline{\hbox to 6.5truein{
\vbox{\hsize=3truein
\psfig{figure=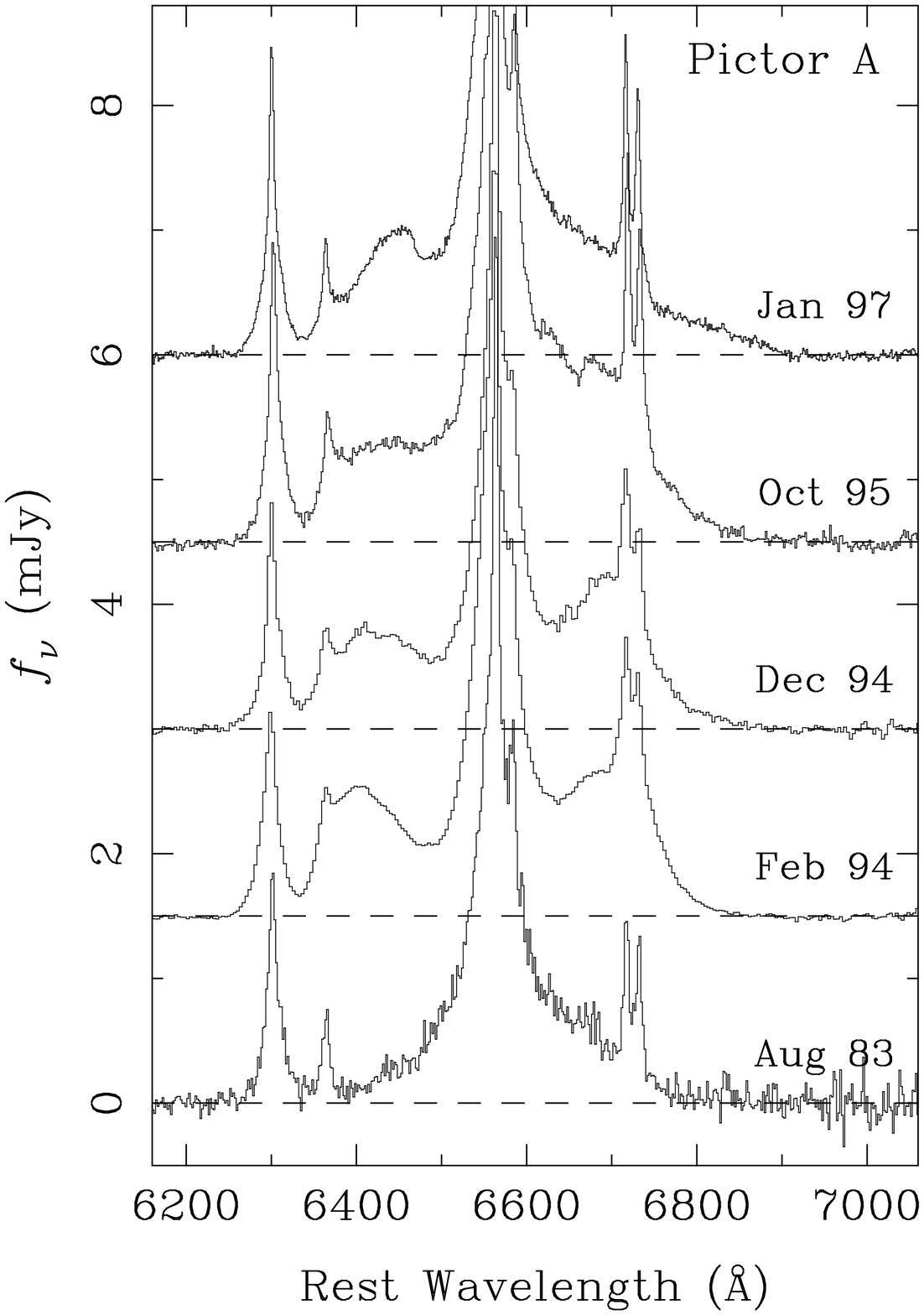,width=3in}
}
\hfill
\vbox to 3 truein {\hsize=3truein
\noindent{\smc Figure~\the\FHa. --} \sml
A compilation of H\a\ spectra of Pictor~A spanning the past 3 years,
along with a spectrum from 1983 showing the H\a\ profile before the
appearance of the displaced peaks. The horizontal dashed lines mark
the zero level of each spectrum. Although the profile has been
varying over the past few years, the displaced peaks were certainly
present at the time of the \asca\ observation in 1996 November.
\vfill}
}}
\bigskip
\endinsert

We have been observing BLRGs with \asca\ with one of our main goals
the investigation of possible differences between the structure of
their inner accretion disks and those of Seyfert galaxies. As such,
our study complements the work on the X-ray spectra and \feka\ lines of
Seyferts by Tanaka et al. (1995), Mushotzky et al. (1995), and Nandra
et al. (1997a). We note parenthetically that the fact that a fair
fraction of BLRGs display superluminal motions is very useful from a
practical perspective because the apparent superluminal speed yields a
stringent upper limit on the inclination of the jet (and hence the
disk) to the line of sight. The inclination of the disk is an
important parameter in the calculation of the model X-ray
spectrum. Another important aim of our program is to use the
properties of the \feka\ lines to test models for the origin of the
double-peaked Balmer lines in BLRGs. The latter goal is closely
related to the former since one possible (and very likely) origin of
the double-peaked Balmer lines is the accretion disk itself. In fact,
the line-emitting disk model of Chen \& Halpern (1989) invokes an ion
torus in the inner disk which has a large large vertical extent and
illuminates the outer disk effectively and drives the Balmer-line
emission. Accordingly, we have targeted BLRGs which are bright X-ray
sources and have double-peaked emission lines. In an earlier paper
(Eracleous, Halpern, \& Livio 1996) we reported the results of our
observation of 3C~390.3, a BLRG with {\it persistent} double-peaked
Balmer lines. We found a resolved \feka\ line in 3C~390.3 with a {\it
FWHM} of 15,000~\kms\ (similar to the width of the Balmer lines) and
most likely coming from an accretion disk at a characteristic radius
of $250\,r_{\rm g}$. The relatively large radius of the line-emitting
region is consistent with the ion torus hypothesis discussed
above. Both the Balmer lines and the \feka\ line could be coming from
the same region in the outer disk.

In this paper we present the X-ray spectrum of the BLRG Pictor~A,
obtained recently with \asca, and we analyze its \rosat\ PSPC spectrum
retrieved from the \rosat\ public archive. Pictor~A's claim to fame is
the abrupt appearance of double-peaked Balmer lines sometime in the
mid-1980s (Halpern \& Eracleous 1994; Sulentic et al. 1995). The
evolution of the H\a\ line profile since the discovery of the
double-peaked lines is depicted in Figure~\the\FHa.  Although the H\a\
line {\it profile} has been varying over the past few years, it has
maintained its overall double-peaked shape and was present at the time
of the \asca\ observation.  Because double-peaked emission lines are
often considered to originate in an accretion disk we thought it
likely that the \feka\ line would have the same profile. The
separation of the displaced peaks in the Balmer lines is of order
14,000~\kms, which is within the resolution of the \asca\ SIS, making
Pictor~A an attractive target. Much to our surprise, however, we do
not detect a strong \feka\ line in Pictor~A at all. To explore the
implications of this result we compare the \feka\ properties of
Pictor~A with those of other BLRGs and Seyfert~1 galaxies observed by
\asca. In \S2 we describe the data and the preliminary reductions,
while in \S3 we compare the observed spectra with models and with the
results of earlier X-ray observations of Pictor~A. In \S4 we place the
X-ray properties of Pictor~A in the context of the X-ray properties of
BLRGs and we compare them with those of Seyferts. We discuss possible
reasons for the weakness of the \feka\ line and the implications for
the structure of the central engines of BLRGs. In \S5 we summarize our
conclusions and present our final speculations. Throughout this paper
we assume a Hubble constant $H_0=50$~km~s\m1~Mpc\m1\ and a
deceleration parameter $q_0=0$.

\bigskip 
\centerline{\smc 2. data and preliminary reductions}
\bigskip

Pictor~A was observed with \asca\ (Tanaka, Inoue, \& Holt 1994) on
1996 November 23--26, with the SIS detectors in {\tt 1-CCD FAINT} mode
and with the GIS detectors in {\tt PH} mode. The data were reduced in
a standard manner, as described by Eracleous et al. (1996); we refer
the reader to this paper for all details associated with the data
reduction as well as the subsequent model fitting. In summary, the
detected photons were screened to eliminate events recorded while the
telescope was pointing very close to the limb of the Earth or events
recorded immediately before or after passage through the South
Atlantic Anomaly or the day/night terminator. Because the original
data were taken in {\tt FAINT} mode, we were also able to correct the
dark-frame error and the echo effect in the SIS detectors.  All of the
above tasks were carried out using the XSELCT/FTOOLS software package
(Blackburn, Greene, \& Pence 1994; Ingham 1994).  Source spectra and
light curves were extracted from a circular (GIS) or rectangular (SIS)
regions centered on the source and large enough to encompass the
entire point-spread function. Background spectra and light curves were
also extracted from annuli around the source. The effective exposure
time and source count rate for each detector are summarized in
Table~\the\Tcounts. The source and background light curves were
inspected for variability on time scales ranging from 15 minutes to
the length of the observation and none was found.

\rosat\ observed Pictor~A with its PSPC twice, first during its All-Sky Survey
(RASS) in the second half of 1990, and then in a pointed observation
on 1991 February 18.  Results from the RASS observation are reported
by Brinkman \& Siebert (1994); we include their reported exposure time
and count rate in Table~\the\Tcounts\ for completeness. We were able
to retrieve the data from the second, pointed observation from the
public archive and use them along with the \asca\ data in our
analysis. As with the \asca\ data we extracted a PSPC source light
curve and spectrum from a circular region of radius 75\asec\ centered
on the source, as well as a background light curve and spectrum from
an annulus around the source. The effective exposure time and count
rate are reported in Table~\the\Tcounts.  Although there was no
discernible source variability within the \rosat\ observation, the
soft X-ray flux measured by the PSPC is approximately a factor of 3
higher then the soft X-ray flux measured by the \asca\ detectors in
the same bandpass. This long-term variability, which comes as no
surprise, is quantified and discussed below.

\topinsert
\centerline{\smc TABLE \the\Tcounts: Exposure Times and Count Rates}
\medskip
\centerline{\vbox{\halign{
# \tabskip 0em & # \hfil \tabskip 1em & \hfil # \hfil \tabskip 2em & 
\hfil # \tabskip 6em  & \hfil # \tabskip 7em & # \hfil \tabskip 3em \cr
\noalign{\hrule \vskip 2pt \hrule \vskip 1em}
& \hw Detector \hw & Year & \hw Exposure Time  \hw & \hw Source Count
Rate$^{\;\rm a}$ \hw  & \hw Background/Source \hw \cr
& & & \hw (s) \hw & \hw (s$^{-1}$) & \cr
\noalign{\vskip 1em \hrule \vskip 1em}
& \asca/\s0                & 1996 & 61,877 & 0.56 & 0.092 \cr
& \asca/\s1                & 1996 & 61,563 & 0.43 & 0.073 \cr
& \asca/\g2\               & 1996 & 68,790 & 0.30 & 0.096 \cr
& \asca/\g3\               & 1996 & 68,762 & 0.36 & 0.105 \cr
& \rosat/PSPC              & 1991 &  4,405 & 0.78 & 0.027 \cr
& \rosat/PSPC$^{\;\rm b}$  & 1990 &    510 & 0.63 & \dots \cr
\noalign{\vskip 1em \hrule \vskip 2pt \hrule \vskip 1em}
}}}
\centerline{\vbox{\hsize=6truein \sml
\item{$^{\rm a}$}
The SIS, GIS, and PSPC count rates refer to the energy ranges 0.6--8.0~keV,
0.9--10.0~keV and 0.1--2.3~keV, respectively.
\item{$^{\rm b}$}
RASS observation (Brinkmann \& Siebert 1994).
}}
\bigskip
\endinsert

\bigskip 
\centerline{\smc 3. comparison of the observed spectrum with models}
\bigskip
\centerline{3.1 \sl The Shape of the Continuum}
\bigskip

The shape of the continuum is best described by a simple power law
modified by interstellar photoelectric absorption\ft{3}. The spectra
of Seyfert galaxies and some BLRGs often include an \feka\ line at a
rest energy of 6.4~keV, which can be modeled (at least at first) by
assuming a Gaussian line profile. In our effort to model the continuum
of Pictor~A we start with the simplest possible model, the power law,
and then we examine whether more components are needed.  We have
investigated whether the data allow for an additional spectral
component at low energies (a soft excess) by adding it to the model
and looking for a significant improvement in the fit.  We found no
evidence for a soft excess regardless of what model we used to
describe it (broken power law, bremsstrahlung, or blackbody).
\footnote{}{\parindent=0pt \item{\ft{3}} \sml Model fits to the continuum were
carried out using the XSPEC software package (Arnaud 1996), adopting
the photoelectric absorption cross-sections of Morrison \& McCammon
1983. We used version 4 (March 3, 1995) of the GIS response matrices,
while for the SIS data we use response matrices created specifically
for this observation, taking into account the evolution of the
detector properties.}

The shape of the X-ray continuum of Seyfert galaxies is observed to
deviate from a simple power law at energies above 10~keV, which has
been interpreted as the result of an additional spectral component,
arising from Compton reflection of the ``primary'' X-rays from cool
dense matter (see \S1). The contribution of the Compton-reflected
X-rays is expected to be small at energies below 10~keV, but we have
searched for it in the GIS spectra, nevertheless. To determine whether
this additional component is present, we fitted the 2.0--10.0~keV GIS
spectra together using a model comprising a power-law spectrum and its
Compton reflection from a slab of cool dense matter (Lightman \& White
1989), which could be the accretion disk. The free parameters of the
model, in addition to the power-law index and normalization, are the
inclination angle of the slab to the line of sight and the solid angle
it subtends to the primary X-ray source. The iron abundance in the
slab was assumed to be solar and the primary continuum was assumed to
be cut off exponentially at 300~keV. The spectrum of reflected X-rays
was computed as a function of inclination angle using the transfer
functions of Magdziarz \& Zdziarski (1995). The result of this
investigation is that a spectral component arising from Compton
reflection is {\it not} required by the data. By varying the
inclination and solid angle of the reflecting slab over the entire
range of allowed values we find that the fit to the data does not
improve significantly, which implies that the two parameters are
formally unconstrained (the significance of the improvement never
exceeds 68\%, or `$1\;\sigma$', for 2 interesting parameters).

\pageinsert
\hbox to 6.5truein {\vsize=4 truein
\psfig{figure=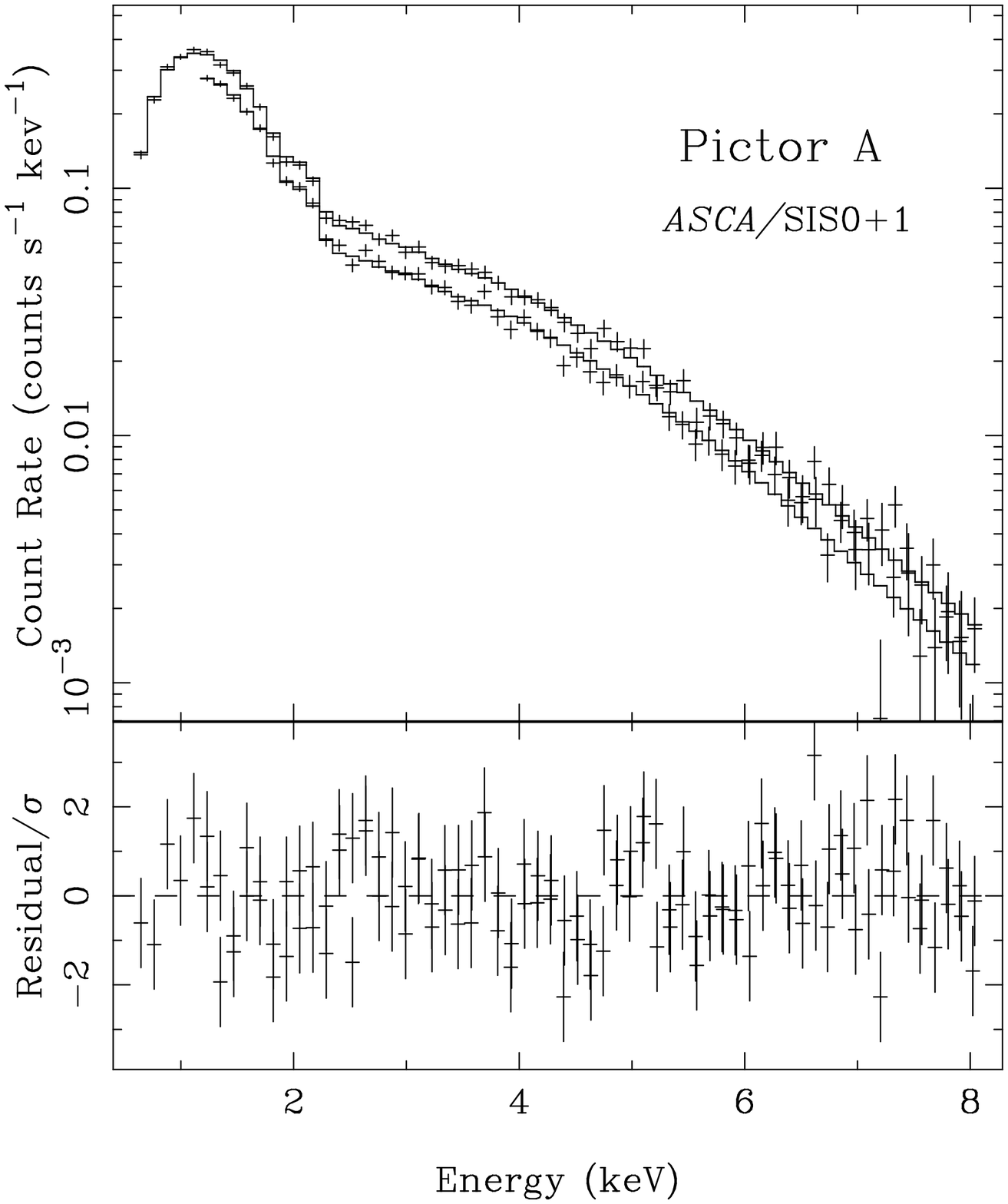,width=3.2in}
\hfill
\psfig{figure=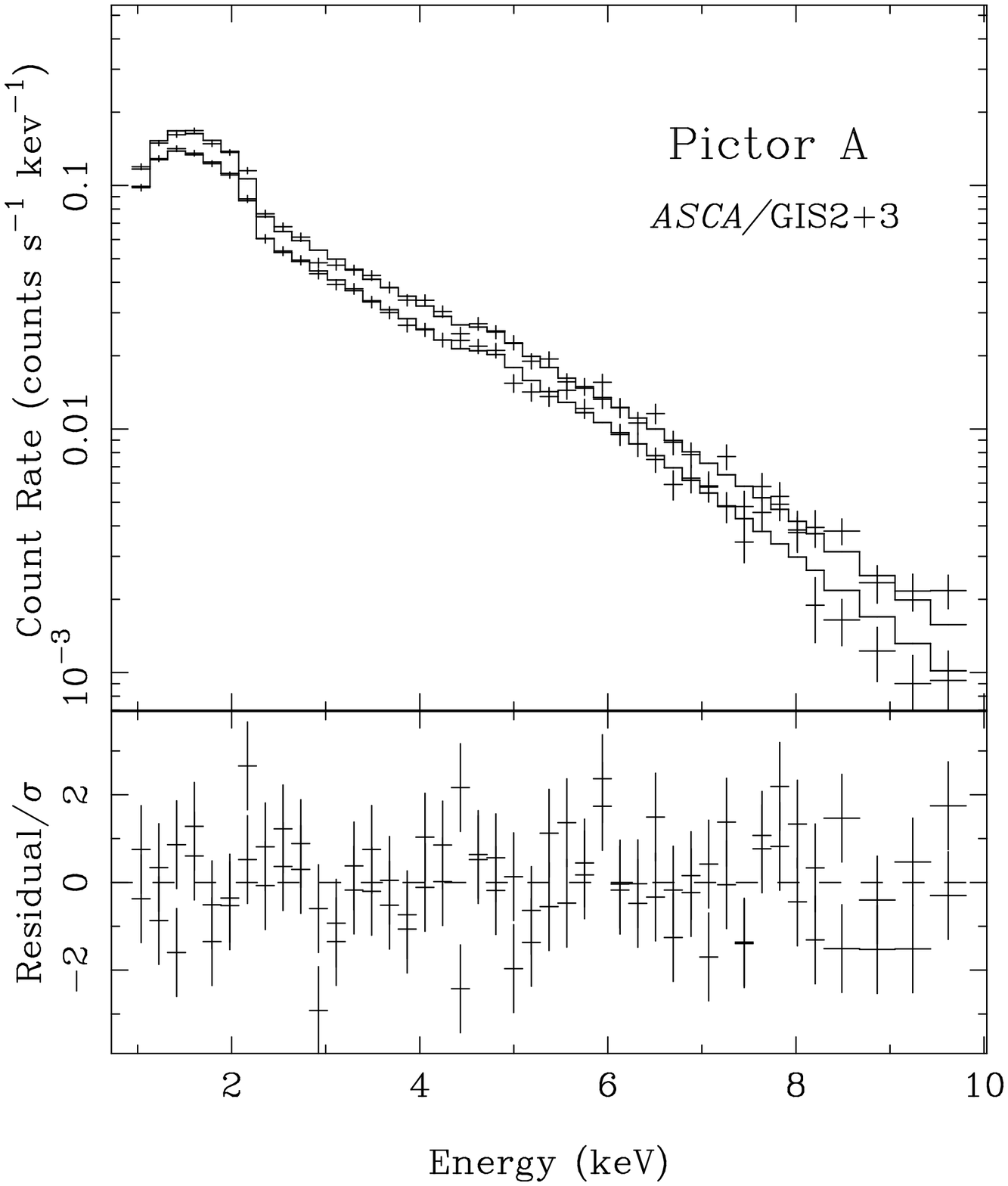,width=3.2in}}
\hbox to 6.5truein {\vsize=4 truein
\psfig{figure=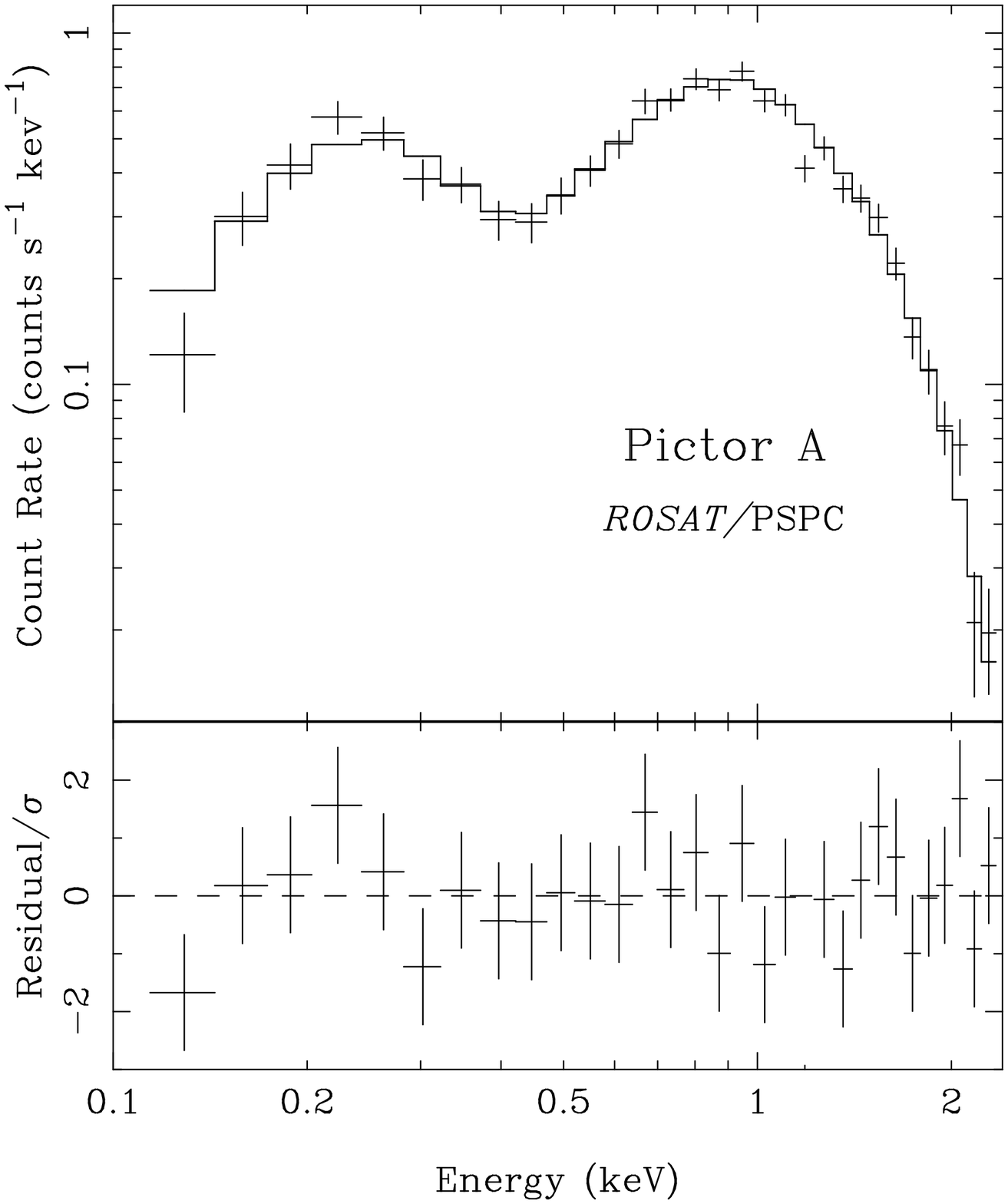,width=3.2in}
\hfill
\vbox to 3 truein {\hsize=3truein
\noindent{\smc Figure~\the\Fspec. --} \sml
The X-ray spectra of Pictor~A from the \asca\ SIS, the \asca\ GIS, and
the \rosat\ PSPC. The top panels show the spectra from each instrument or
pair of instruments with the best-fitting power-law model(s) overlayed for
comparison. The lower panels show the post-fit residuals  scaled by the error
bar at each point. 
\vfill}}
\vfill
\endinsert

By fitting the observed distribution of counts per energy channel from
each instrument separately we derive model parameters which are
consistent with each other, and thus we justify carrying out joint
fits to the data from the two SIS detectors and similarly to the data
from the two GIS detectors. We find a discrepancy between the model
and the \s1\ data at energies between 0.5 and 1~keV. By ignoring this
region of the \s1\ spectrum we obtain a good fit with consistent
parameters to data from all instruments individually, and to data from
pairs of instruments fitted simultaneously.  Therefore, we adopted
this empirical remedy. We note in defense of our approach that Yaqoob
(1996) finds the same systematic discrepancy that we do using two
separate observations of 3C~273 taken during the performance
verification phase and during cycle 4. The parameters describing the
best-fitting model are summarized in Table~\the\Tfit.  The weighted
mean power-law index determined from the SIS and GIS spectra is
$1.77\pm 0.03$ and the equivalent hydrogen column density of of
absorbing matter is ($6\pm2$)\tten{20}~cm\m2, as measured by the SIS
detectors, which are the most sensitive at low energies (all error
bars correspond to the 99\% confidence limits for 2 interesting
parameters).  The spectra of counts per energy channel from the GIS
and SIS detectors are shown in Figure~\the\Fspec$a,b$. The model
spectra, after convolution with the telescope and detector response
matrices are overlayed for comparison. The uncertainties in the
best-fitting model parameters can also be expressed graphically in
terms of the 99\% confidence contours in a 2-parameter plane, as shown
in Figure~\the\Fcont. The GIS and SIS confidence contours overlap at
the 90\% and 99\% confidence levels indicating that the results from
the two sets of detectors are consistent with each other.

\pageinsert
\centerline{\smc TABLE~\the\Tfit: Best-Fitting Model Parameters and Fluxes}
\medskip
\centerline{\vbox{\halign{
# \hfil &  \hfil # \hfil \tabskip 0em & # \hfil & # \hfil & \hfil # \hfil
\tabskip 1em & \hfil # \hfil \tabskip 1em & \hfil # \hfil \tabskip 0em\cr
\noalign{\hrule\vskip 2pt \hrule \vskip 6pt}
\hw Instrument \hw &  Year & \hw $\chi_{\nu}^2$/d.o.f. \hw & 
\hw Photon \phantom{$^{\;\rm a}$} \hw & Column \phantom{$^{\;\rm a}$}  &  
\multispan2{\hw Flux$^{\;\rm b}~(10^{-11}~{\rm erg~cm^{-2}~s^{-2}})$\hw} \cr
& & &  \hw Index$^{\;\rm a}$ \hw &  Density $^{\;\rm a}$ &
\multispan2{\hrulefill} \cr
& & & & (\ten{20}~cm\m2) & 0.7--2.4~keV & 2--10 keV \cr
\noalign{\vskip 6pt \hrule \vskip 6pt}
\noalign{\vskip 6pt} 
\multispan7{\hw Measurements from This Paper \hw} \cr
\noalign{\vskip 6pt}
\asca/\s{0+1} & 1996 & 1.066/106 & $1.77^{+0.04}_{-0.05}$
              & $6\pm2$   & $0.77\pm0.03$ & $1.4\pm0.1$ \cr
\asca/\g{2+3} &  1996 & 1.254/82  & $1.76^{+0.06}_{-0.05}$
              & $8\pm4$   & \dots & $1.5\pm0.2$  \cr
\rosat/PSPC   & 1991 & 0.832/26  & $1.7^{+0.3}_{-0.2}$
              & $4.4^{+1.6}_{-0.9}$ & $1.8\pm0.2$ & \dots \cr 
\noalign{\vskip 6pt} 
\multispan7{\hw Measurements from the Literature \hw} \cr
\noalign{\vskip 6pt}
\rosat/PSPC   & 1990$^{\;\rm c}$ & \hw \dots \hw & $2.3\pm0.6$ & $7\pm3$ 
              & $1.94^{+0.12}_{-0.06}$ & \dots \cr
\exosat/ME+LE & 1986$^{\;\rm d}$ & \hw \dots \hw & $1.8\pm0.3$ & $7^{+7}_{-4}$ 
              & $1.03\pm0.06$ & $1.73\pm0.10$ \cr
\einstein/IPC & 1980$^{\;\rm d}$ & \hw \dots \hw & $1.5^{+0.5}_{-0.3}$ & $6^{+14}_{-4}$
              & $0.71\pm0.03$ & \dots \cr
\noalign{\vskip 6pt \hrule\vskip 2pt \hrule\vskip 1em}
}}}
\centerline{\vbox{\hsize=6truein \sml
\item{$^{\rm a}$} 
The uncertainty in the model parameters corresponds to the 99\% confidence 
limits for 2 interesting parameters. 
\item{$^{\rm b}$} 
The quoted flux has been corrected for interstellar photoelectric absorption. 
The error bars reflect the uncertainties in both the spectral index and the
normalization.
\item{$^{\rm c}$} 
RASS observation reported by Brinkmann \& Siebert (1994)
\item{$^{\rm d}$}
Singh et al. (1990). The \einstein\ IPC flux has also been published
by Kruper, Urry, \& Canizares 1990.
}}

\centerline{\psfig{figure=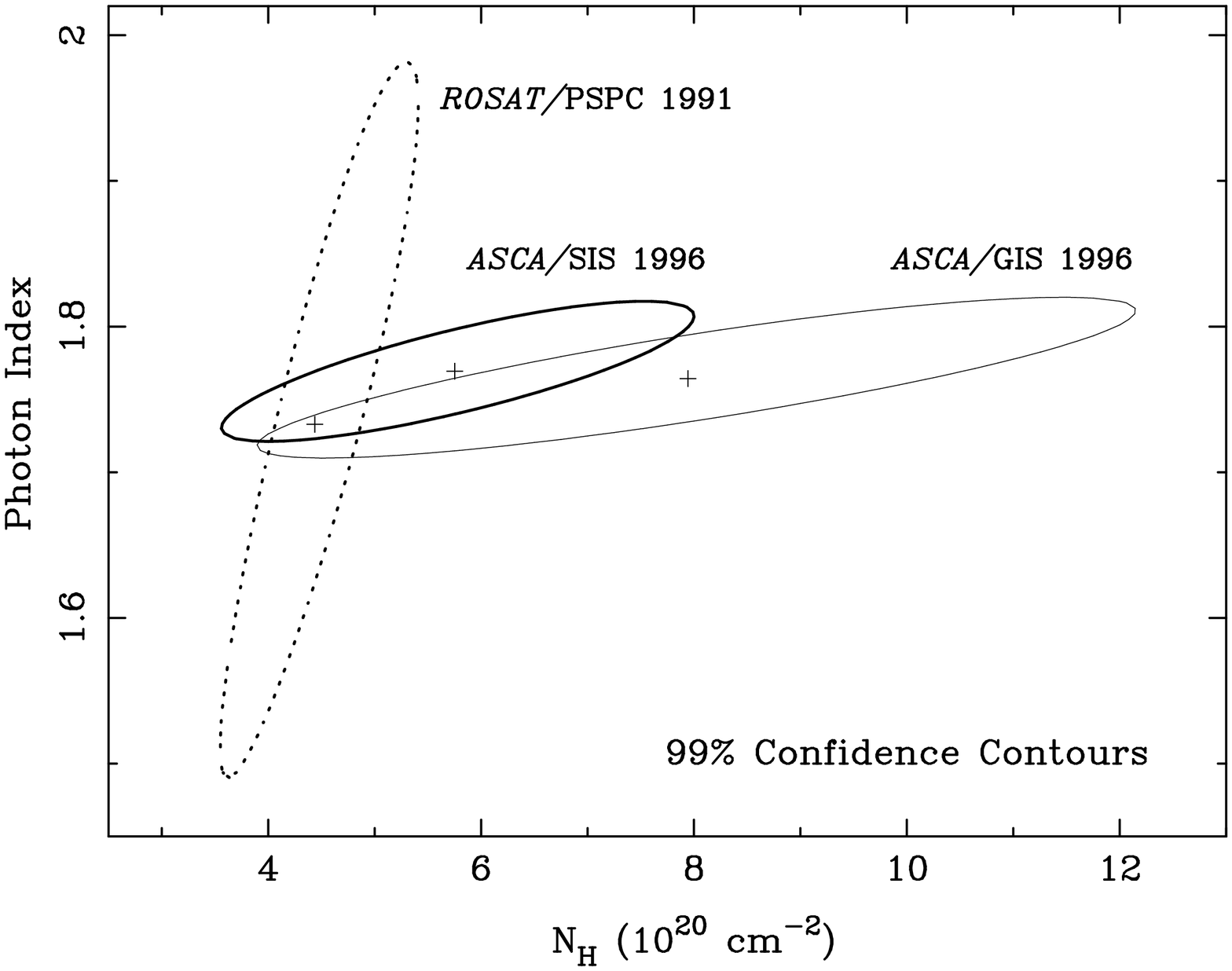,height=3.5in,angle=-90}}
\vfill
\centerline{\vbox{\noindent{\smc Figure~\the\Fcont. --} \sml
The 99\% confidence contours showing the correlated uncertainties between the
power-law index and the column density of the best-fitting model for each
instrument. The cross near the center of each contour marks the best fit
obtained for that instrument.}}
\endinsert

The \rosat\ PSPC spectrum was fitted separately using the same model
as the one used for the \asca\ spectra. The spectrum and best-fitting
model are shown in Figure~\the\Fspec$c$, while the best-fitting model
parameters and their uncertainties are included in Table~\the\Tfit.
The model that describes the \rosat\ spectrum is consistent, within
uncertainties, with the model describing the \asca\ spectra, save for
the normalization. This is illustrated in Figure~\the\Fcont, where the
99\% confidence contour for the parameters of the \rosat\ spectrum is
compared to the corresponding contours deprived for the \asca\
spectra. The difference in model normalizations between the \rosat\
and \asca\ spectra is reflected in the 0.7--2.4~keV fluxes reported in
Table~\the\Tfit\ (this band was chosen because it is common to the
\rosat\ PSPC and the \asca\ SIS). The flux appears to have increased
by a factor of 2.6 between 1991 and 1996, which is most likely the
result of intrinsic variability of the source. The spectral index and
the absorbing column density do not appear to have changed within the
uncertainties.

We have compared our measurements with published reports of the
spectrum of Pictor~A measured in the 1980s with the \einstein/IPC and
\exosat/ME+LE by Singh, Rao, \& Vahia (1990). The results of these
early observations are included in the second part of
Table~\the\Tfit. A simple power-law model provides the best
description of the \exosat\ and \einstein\ spectra, just as it does
for the \asca\ and \rosat\ spectra.  Also included in Table~\the\Tfit\
are the spectral parameters measured by Brinkmann \& Siebert (1994)
from the RASS data.  The soft (0.7--2.4~keV) X-ray flux varies by less
than a factor of 3 over all observations of Pictor~A to date. The
spectral shape and absorbing column density do not appear to have
changed (within uncertainties, of course). Since the X-ray spectrum
between 0.5 and 10 keV is described well by a power law, it is
reasonable to conclude that the 2--10~keV flux follows the variations
of the soft X-ray flux. Finally, we note that the Galactic hydrogen
column in the direction of Pictor~A is 4.2\tten{20}~cm\m2\ (Heiles \&
Cleary 1979). This is consistent, within errors, with all measurements
of the absorption from the X-ray spectra, suggesting that the
interstellar medium of the Galaxy is responsible for most of the
observed absorption.

\goodbreak
\bigskip
\centerline{3.2 \sl The {\rm \feka} Line: Too Weak to Be Measurable}
\bigskip

A ubiquitous feature in the \asca\ X-ray spectra of Seyfert galaxies is an
\feka\ line at a rest energy of 6.4~keV (see \S1). This line is thought  to
arise by fluorescence in the same cool, dense matter where Compton
reflection of the continuum takes place. Our preliminary inspection of
the spectrum showed no evidence for a line. Hence, to investigate
whether a weak line is present and to set limits on it equivalent
width (hereafter $EW$) we fitted the \asca\ spectra in the range
3.0--8.0~keV with a model consisting of a power-law continuum and a
line. Photoelectric absorption was not included because its effects
are negligible at column densities as low as what was measured in
\S3.1.  We carried out the exercise under two different assumptions
for the profile of the line. The first assumption is that the line has
a Gaussian profile. The full velocity width of the line at half
maximum, in its rest frame, is given by $FWHM=\sqrt{8\,\ln2}\; c
\;\sigma_{\rm obs}/E_{\rm obs}$, where $E_{\rm obs}$ and $\sigma_{\rm
obs}$ are the observed energy and energy dispersion of the line.  The
second assumption is that the line profile is disk-like, computed
according to the model of Fabian et al. (1989). In this model the line
originates in an annulus of a disk which is inclined to the line of
sight and whose emissivity varies with radius as a power
law. Motivated by the properties of \feka\ line in Seyfert galaxies,
we have fixed the inner and outer radius of the line-emitting part of
the disk to $6\, r_{\rm g}$ and $1000\,r_{\rm g}$ and set the
emissivity proportional to $r^{-3}$ in our implementation of this
model. Thus, the inclination angle of the disk, which is a free
parameter, effectively controls the width of the line (roughly as {\it
FWHM} $\propto\sin\, i$, where $i$ is the disk inclination). The main
difference between the two model profiles is that the Gaussian profile
is symmetric, while the disk-like profile is skewed with the red wing
more prominent than the blue.

\pageinsert
\centerline{\psfig{figure=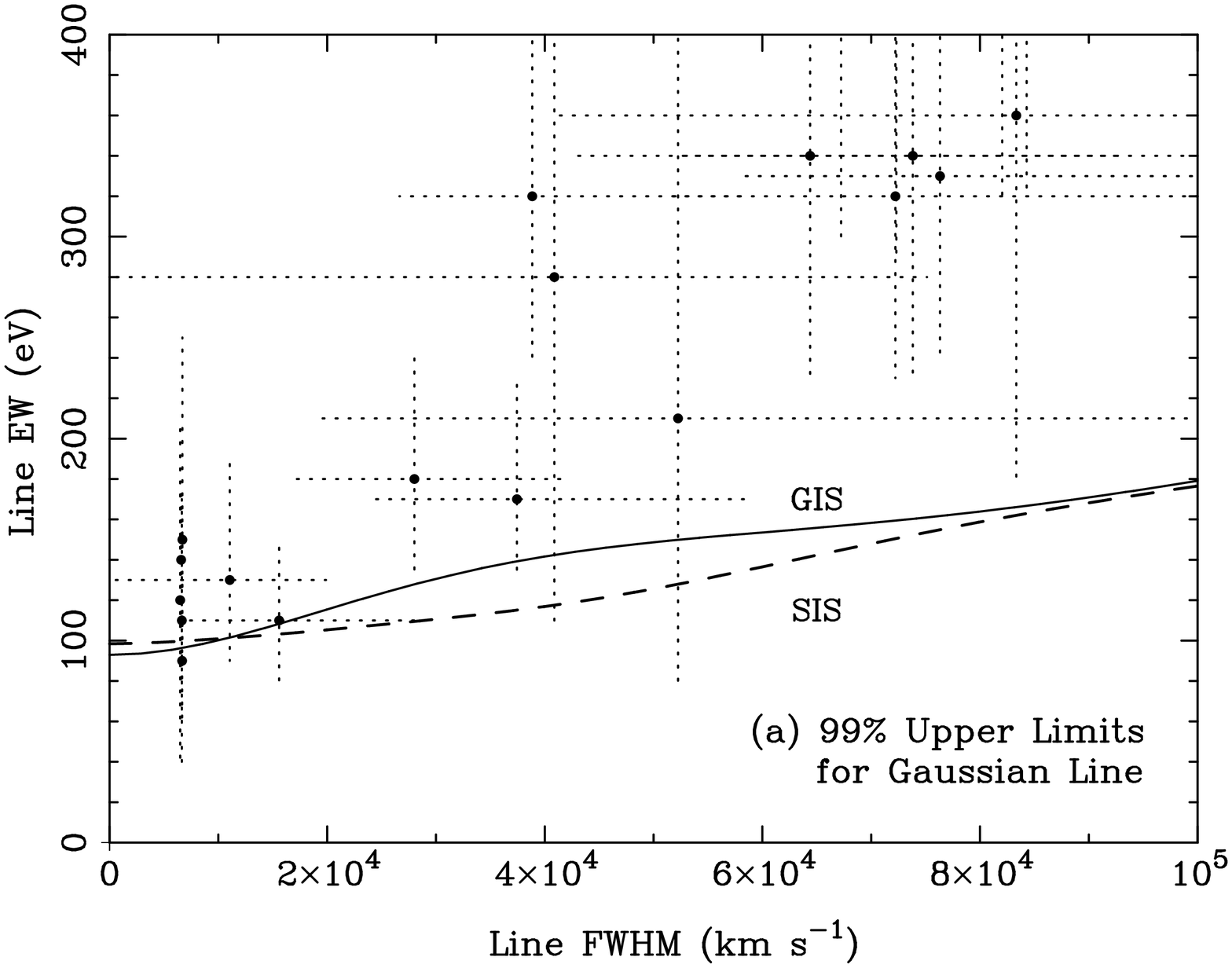,height=3.4in,angle=-90}}
\centerline{\psfig{figure=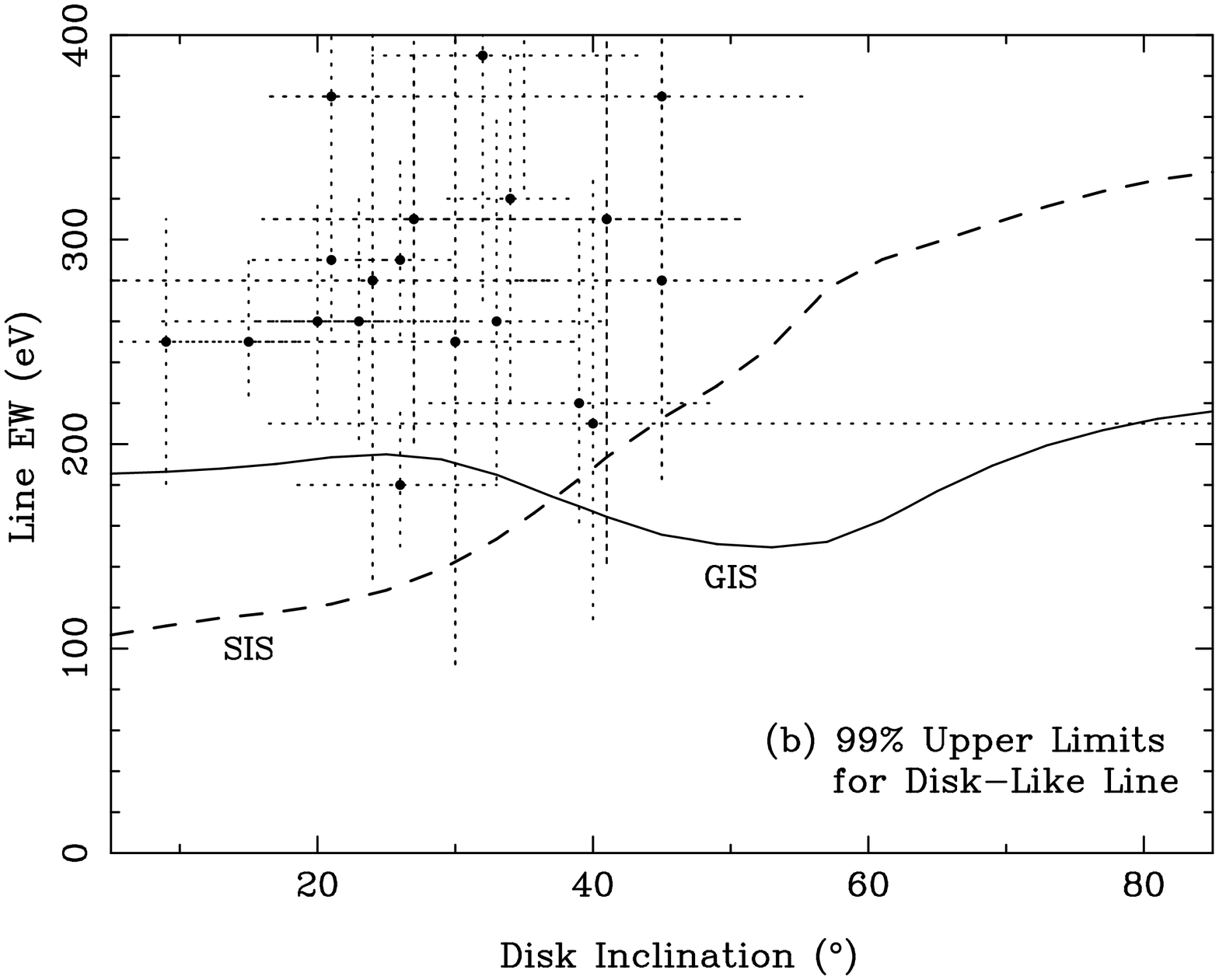,height=3.4in,angle=-90}}
\vfill
\centerline{\vbox{\hsize=6.5truein \noindent{\smc Figure~\the\Flimits. --} \sml
(a) The upper limit to the \feka\ $EW$ (measured in the observer's frame) as a
function of the {\it FWHM} of the Gaussian profile assumed for the line.  For
reference, we note that the resolution of the SIS  corresponds to a {\it FWHM}
of 12,700~\kms, and that of the GIS to 22,800~\kms. (b) The upper limit to the
\feka\ $EW$ (measured in the observer's frame) as a function of disk inclination
in the case where the line is assumed to have a disk-like profile. The
inclination of the disk effectively controls the width of the line. In each
panel we also plot the $EW$s and velocity widths measured in individual
Seyfert galaxies by Nandra et al. (1997$a$) for comparison. Error bars for each
measurement are also plotted as dotted lines in order to make the comparison
as fair as possible. 
}}
\endinsert

We applied each of our assumed models to the data from each pair of
instruments.  The strategy we adopted was to fit the continuum to the data in
two windows straddling the line (3.0--4.0~keV and 7.0--8.0~keV; at $z=0.035$,
the line is expected to appear at 6.18~keV) and freeze it. The two continuum
windows were chosen in view of the fact that the \feka\ lines observed in the
\asca\ spectra of Seyfert galaxies are extremely broad, with extended red
wings.  After setting the continuum level, the line was added to the
model which was fitted to the entire 3.0--8.0~keV range, allowing the line
flux and width to vary freely. To determine the upper limit to the $EW$ of the
line allowed by the data we scanned the line width--intensity parameter plane
(or inclination-intensity plane, in the case of the disk-like profile model)
looking for sets of parameters that yielded a significantly better fit to the
data compared to the simple continuum model. The result of this parameter-space
search is a set of upper limits on the flux of line photons as a function of
the line width or the inclination of the disk. Because the continuum was held
fixed as the line parameters were varied, the flux of the line can be
translated to an $EW$ by a simple scaling. Therefore, we present our results in
the form of upper limits on the line $EW$ versus line width or disk inclination
in Figure~\the\Flimits. In this form, the upper limits that we determine
can be compared directly with the measured $EW$s of lines in Seyfert galaxies
from the compilation of Nandra et al. (1997$a$). To this end we plot in
Figure~\the\Flimits\ the individual $EW$s measured in Seyferts by Nandra et al
(1997$a$) {\it along with their error bars}. We note that  the method we have
used here to determine the $EW$ upper limits is not necessarily the same as
that used by other authors to determine error bars in the $EW$ of {\it
detected} lines. The difference lies in that we have fixed the continuum level
and allowed only the line parameters to vary freely. It is possible to allow
the continuum level to be free as well, as other authors may have done, which
would make the error bars on the line $EW$ somewhat larger. Since we are
interested in comparing the $EW$ {\it upper} limits deprived here with the 
{\it lower} $EW$ error bars deprived for Seyfert galaxies, the difference in
methodology will not affect our conclusions.

The upper limit (at 99\% confidence for 2 interesting parameters) to
the observed $EW$ of a line which is unresolved by the SIS (i.e., {\it
FWHM}$<1.28\times 10^4$~\kms) is 100~eV. For a resolved Gaussian line
(either in the SIS or the GIS) the $EW$ upper limit is a function of
the width of the line. This is a manifestation of the fact that for a
fixed $EW$ a broader line can hide in the noise more easily than a
narrower one. Moreover, the minimum detectable $EW$ for a fixed
exposure time with any given instrument is determined by the
competition between its effective area and its spectral resolution. At
the off-axis angles at which Pictor~A was observed the effective area
of \g3 at 6.2~keV is 25\% higher than that of \s0. Therefore, the
minimum detectable $EW$ in the GIS is comparable to that in the SIS
even though the spectral resolution of the former is almost a factor of
2 poorer than that of the latter (the exposure time in the GIS is also
about 12\% longer than in the SIS). For a Gaussian line that is as
broad as the lines detected in Seyferts by Nandra et al. (1997$a$)
the $EW$ upper limit is 135~eV, assuming that {\it FWHM}$\approx
6\times 10^4$~\kms. This is below the range of $EW$ measured in
Seyferts under the same assumptions, as shown in
Figure~\the\Flimits$a$.  If the line is assumed to have a disk-like
profile, the $EW$ upper limit depends on the assumed inclination of
the disk since this determines the width of the
line. Figure~\the\Flimits$b$ shows the relation between the the $EW$
upper limit and the disk inclination for both the SIS and the GIS. If
the disk inclination is assumed to be comparable to that measured for
Seyferts by fitting their line profiles, the implied $EW$ upper limit
for Pictor~A is 180~eV, which is well below the range of Seyfert
galaxy $EW$s. We note that for assumed disk inclinations above 35\deg\
the resulting model line profiles are so broad that the higher
sensitivity of the GIS compared to the SIS outweighs its poor spectral
resolution and makes the $EW$ limits from the former instrument more
stringent than those from the latter.

\bigskip 
\centerline{\smc 4. discussion}
\bigskip

\centerline{4.1. \sl Comparison With Other BLRGs and With Seyferts}
\bigskip

The absence of an \feka\ line from the spectrum of Pictor~A is
surprising because such lines are detected in the spectra of almost
all Seyfert~1 galaxies and BLRGs observed by \asca\ (Nandra et
al. 1997$a$; see also Table~\the\TBLRGs). We note that Singh et
al. (1990) found a very strong \feka\ line in the \exosat\ spectrum of
Pictor~A ($EW\sim 1$~keV) but those data could well be affected by
systematic errors associated with the background at energies higher
than 6~keV.  To explore the implications of the weakness of the line
we compare our limits on the line properties with the properties of
lines detected in other BLRGs and in Seyferts.  We also investigate
whether the weakness of the line can be understood in the context of
the known anticorrelation between X-ray luminosity and \feka\ $EW$
(Nandra et al. 1997$b$). To this end we collect in Table~\the\TBLRGs\
information of the properties of \feka\ lines of BLRGs from the
literature. We include in this table constraints on the the
inclination angle of the jet (and hence the disk) of each object
derived from its radio properties. Namely, (a) we obtain a {\it lower}
limit on the jet inclination by requiring that the size of the
double-lobed radio source not exceed the largest sizes observed in
radio galaxies, and (b) for three objects displaying superluminal
motion we obtain an {\it upper} limit on the jet inclination from the
measured superluminal speed (see the detailed discussion of the method
by Eracleous et al. 1996). In Figure~\the\FBLRGs\ we plot the \feka\
rest-frame $EW$s of BLRGs against luminosity, and we also indicate the
region of this diagram occupied by Seyferts, according to Nandra et
al. (1997$b$), for comparison (the width of the grey band in this
figure corresponds to the dispersion in the $EW$ of objects in the
same luminosity bin).

\topinsert
\centerline{\smc TABLE~\the\TBLRGs: BLRG Iron Line Properties}
\medskip
\centerline{\vbox{\halign{
# \hfil \tabskip 0.5em & \hfil # \hfil \tabskip 0.5em & \hfil # \hfil \tabskip 0em 
& \hfil # \hfil \tabskip 1em & \hfil # \hfil \tabskip 0em &  \hfil # \tabskip 1em 
& 
# \hfil \tabskip 1em \cr
\noalign{\hrule\vskip 2pt \hrule \vskip 6pt}
\hw Object \hw & $z$ & $L_{\;\rm x\;}$(2--10 keV) & Rest $EW$ & {\it FWHM} & 
\hw Inclination \hw & \hw References$^{\;\rm a}$ \hw \cr
& & (erg s\m1) & (eV) & ($10^3$~\kms) & \cr
\noalign{\vskip 6pt \hrule \vskip 6pt}
3C 109   & 0.306 & 2.1\tten{45} & $390^{+780}_{-260}$ & $90^{+112}_{-50}$  & 
$i>35^{\circ}$ & 1,2 \cr
3C 111$^{\rm b}$& 0.048 & 4.1\tten{44} & $<130$              & \hw\dots\hw        
& $37^{\circ}>i>24^{\circ}$ & 2,3,4,5 \cr
3C 120$^{\rm c}$& 0.033 & 2.0\tten{44} & $380^{+100}_{-140}$ & $91^{+34}_{-35}$   
& $14^{\circ}>i>1^{\circ}$ & 6,7,8 \cr 
Pictor A & 0.035 & 7.9\tten{43} & $<140$          & \hw\dots\hw        & 
$i>24^{\circ}$ & 2,9,10 \cr
3C 382   & 0.059 & 6.0\tten{44} & $950^{+1100}_{-320}$ & $197^{+109}_{-55}$ & 
$i>15^{\circ}$ & 2,11 \cr
3C 390.3 & 0.056 & 2.0\tten{44} & $190^{+100}_{-70}$  & $15^{+15}_{-8}$    & 
$33^{\circ}>i>19^{\circ}$ & 2,12,13 \cr
3C 445   & 0.057 & 8.5\tten{43} & $270^{+130}_{-90}$  & $18^{+18}_{-14}$   & 
$i>60^{\circ}$ & 14,15 \cr 
\noalign{\vskip 6pt \hrule\vskip 2pt \hrule\vskip 1em}
}}}
\medskip\centerline{\vbox{\hsize=6.2truein \sml
\item{$^{\rm a}$} 
{\sl References:} 
(1) Allen et al. 1997; (2) Nilsson et al. 1993; (3) Reynolds et al. 1997; (4)
Eracleous \& Halpern 1997$b$; (5) Vermuelen \& Cohen 1994; (6) Grandi et al.
1997; (7) Balick, Heckman \& Crane 1982; (8) Zensus 1989; (9) this work; (10)
Jones \& McAdam 1992; (11) Reynolds 1997; (12) Eracleous et al. 1996; (13) Alef
et al. 1994; (14) Sambruna et al. 1997;   (15) McCarthy, van Breugel, \& Kapahi
1991.
\item{$^{\rm b}$} 
Reynolds et al. (1997) report a marginal detection of an \feka\ line with an
$EW\approx 100$~eV under the assumption of a disk-like profile. Our own
independent analysis of the same data (Eracleous \& Halpern 1998$b$) yielded an
upper limit to the $EW$ of a Gaussian line of 130~eV, which we adopt here.  
\item{$^{\rm c}$} 
The upper limit on the inclination of the jet in 3C~120, $i<$14$^{\circ}$, is 
based on a superluminal speed of $\beta_{\rm app}$=8.1, reported by Zensus
(1989). Other authors (Wehrle et al. 1992; Walker, Walker, \& Benson 1988)
report lower superluminal speeds ($\beta_{\rm app}=4.1$ and $\beta_{\rm app}$
=3.7$\pm$1.2, respectively), which imply $i\ls$25$^{\circ}$.
}}
\bigskip
\endinsert

\topinsert
\centerline{\psfig{figure=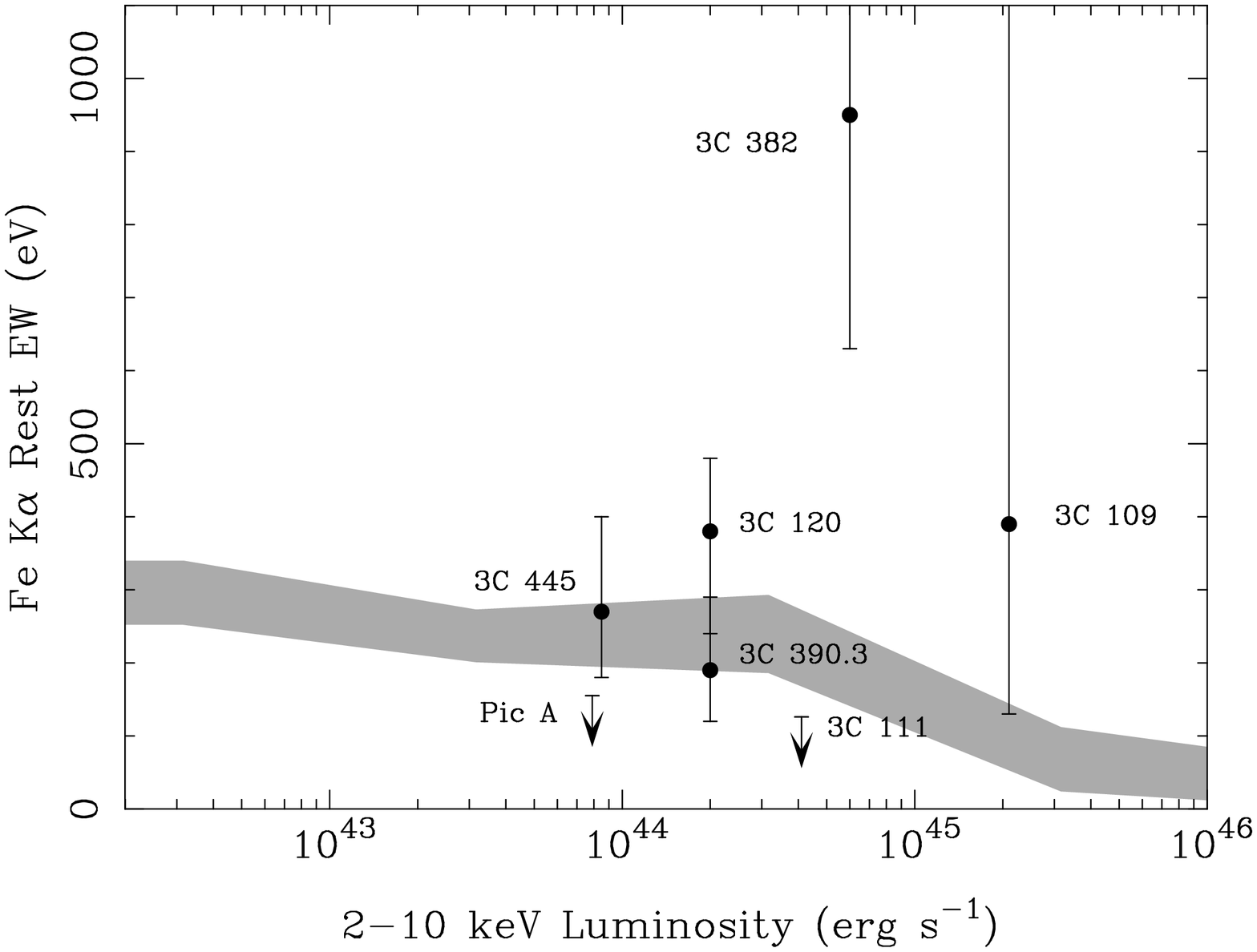,height=3.5in,rheight=4in,angle=-90}}
\medskip \centerline{\vbox{\hsize=6 truein
\noindent{\smc Figure~\the\FBLRGs. --} \sml
The rest-frame $EW$s of the \feka\ lines of BLRGs plotted against
luminosity.  The data themselves are given in Table~\the\TBLRGs. The
shaded band shows the region of the $EW$-luminosity plane occupied by
Seyfert galaxies according to Nandra et al. 1997$b$. The width of this
band reflects the dispersion in $EW$s among objects in the same
luminosity bin. In the highest luminosity bin the sample of Nandra et
al. (1997$b$) includes a significant number of radio-loud objects.}}
\bigskip
\endinsert

Figure~\the\FBLRGs\ has a number of interesting and noteworthy
features. First, with the exception of 3C~109 (which is best
classified as a quasar rather than a BLRG because of its high
luminosity) all objects cluster in a narrow luminosity range around
\ten{44}~erg~s\m1. Second, there is a large scatter in $EW$s among
BLRGs, ranging from $<130$~eV (3C~111) to $>600$~eV (3C~382). Third,
the large $EW$s observed in 3C~120 and 3C~382 are particularly
disturbing (or interesting, depending on one's perspective) if they
are considered in combination with the line widths ({\it FWHM}). As
the authors who report the original results point out (Reynolds 1997;
Grandi et al. 1997) the lines are so broad that it is difficult to
understand their widths in the context of a line-emitting disk model
for their origin. In the case of 3C~120, the disk is thought to be
viewed almost face-on ($i<14^{\circ}$) and yet its \feka\ line has an
{\it observed} {\it FWHM} of 91,000~\kms. If this is true, then the
intrinsic (deprojected) velocities of the line wings should exceed the
speed of light. In the case of 3C~382 the {\it observed} {\it FWHM}
corresponds to half the speed of light, implying that the wings of the
line should correspond to superluminal speeds regardless of the
inclination. We note that such large widths and $EW$s could have come
about if the continuum level was underestimated in the original
analysis of the data.  Such a problem can come about if, for example,
the continuum level is allowed to vary freely when fitting the
line. Under these conditions a very broad line can mimic the continuum
and result in an erroneous estimate of the equivalent width. Indeed,
Wo\'zniac et al. (1998) find from their independent analysis of the
same data that the lines are weaker and narrower than the original
reports suggest.  However, since the results of Wo\'zniac et
al. (1998) are not yet published, we take the reported line properties
at face value, but we regard them with great caution. The moral of
this comparison is that the orientation of the disk (or the jet) does
not seem to play an important role in determining the properties of
the \feka\ line. We reach this conclusion by considering the fact that
both 3C~390.3 and 3C~445 which have very different jet inclinations
have relatively weak and narrow lines, while 3C~382 and 3C~120 also
have very different inclinations and yet have very strong lines and
very broad. With this observation in mind we turn our attention to
possible reasons for the weakness of the
\feka\ line in Pictor~A, which we discuss in the next section.

\bigskip 
\centerline{4.2. \sl What Hapenned to the {\rm \feka} Line?}
\bigskip 

Production of the fluorescent Fe K$\alpha$ line requires that an X-ray 
continuum illuminates cold target material of sufficient optical depth and
covering fraction.  In ordinary Seyfert galaxies, it is thought that the only
target with the necessary properties is the accretion disk. In the context of
this scenario we consider possible reasons for the weakness of the Fe~K$\alpha$
line in Pictor~A. We seek an explanation which is not just specific to Pictor~A,
but rather one that is also consistent with all (or at least most) of the
available data as they are summarized in Table~\the\TBLRGs, including the
widths of the \feka\ lines and the independent information on the disk
orientation obtained from the radio properties.

\item{(a)} {\it A Highly Inclined Disk:} 
The inclination angle of the accretion disk of Pictor~A is likely to
be large.  The radio core-to-lobe luminosity ratio ($\log R=-1.82$;
Jones \& McAdam 1992) and the projected separation of the radio lobes
(Table~\the\TBLRGs) imply $i > 24^{\circ}$.  If this is the case, then
there are two effects which conspire to hide the line.  First, the
$EW$ of the line predicted by models is a function of the disk
inclination; at $i=50^{\circ}$ it is 30\% smaller than the face-on
value (George \& Fabian 1991; Matt et al. 1991). Second, the line
appears broader if the disk is viewed at a higher inclination, and
hence for a fixed line flux the $EW$ threshold for detecting it
increases ($EW_{\rm min} \propto FWHM \propto \sin i$). This
intepretation is consistent with the lack of a Compton-reflected
spectral component at the highest observable energies. However, the
same explanation {\it cannot} be invoked for 3C~111 or 3C~390.3 whose
superluminal motion and projected radio lobe separation require that
their disks are viewed close to face on (Table~\the\TBLRGs). Moreover,
3C~109 does not fit into the picture either because it has a strong
line although the inclination of its disk appears to be large.

\item{(b)} {\it Inner Disk Structure Different from Seyferts:}
Production of the large equivalent widths in Seyfert galaxies is
difficult to achieve in an accretion disk geometry unless the covering
fraction is close to 50\%. This requires the hard X-ray source to
``hug'' the accretion disk, as would a geometrically thin
corona. BLRGs could have a different accretion disk structure, such as
a spherical corona above a thin disk, or an ion torus that results
from an instability in the inner disk (within a few hundred $r_{\rm
g}$).  In the latter case, the covering fraction of the cold outer
disk as seen by the X-ray source is at most 25\%, and possibly
considerably less than that, as shown by simple geometrical
considerations (Chen \& Halpern 1989). Moveover, the Keplerian
velocity in the line-emitting part of the disk is only around
10,000--30,000~\kms, with the consequence that the observed lines in
BLRGs should be narrower than in Seyferts. This idea is analogous to
the scenario put forth by Wo\'zniac et al. (1998) in which the \feka\
lines were produced in medium far away from the central black hole
whose column density was not high enough to give rise to Compton
reflection. Here, however, we make the specific suggestion that the
medium where the lines originate is the outer accretion disk
itself. This explanation could also apply to 3C~390.3, whose
Fe~K$\alpha$ line is neither as broad nor as strong as those of
Seyferts, and prehaps also to 3C~445. However, the same explanation
would not apply to objects such as 3C~109, 3C~120, or 3C~382 which
seem to have strong, broad Fe~K$\alpha$ lines.

\item{(c)} {\it Continuum Beamed Away from the Disk:} 
The observed continuum could be beamed in a direction away from the
accretion disk, the latter being the source of Fe~K$\alpha$ lines in
Seyferts. The result is that the disks of BLRGs are not illuminated
effectively and hence their lines are weak.  However, they may still
have weak narrow Fe~K$\alpha$ lines if the continuum illuminates part
of the broad-line region (Yaqoob et al. 1993).  But this would appear
to be a rather contrived explanation because it applies only to
Pictor~A and 3C~111. Most BLRGs observed with \asca\ do have broad
\feka\  lines (some are quite strong) and there is no obvious correspondence 
between their continuum luminosity and the orientation of their disks; hence
they do not support this hypothesis.

\item{(d)}{\it Continuum Beamed Towards the Observer:} 
If the observed continuum is beamed towards the observer (presumably
because the source is associated with the jet), then the continuum
level is boosted and the observed $EW$ of an unbeamed line
decreases. However, the jet orientation arguments presented in (a)
above make this an unlikely explanation for Pictor~A.  For an assumed
jet Lorentz factor of 10, the only object in Table~\the\TBLRGs\ for
which beaming could boost the continuum flux by a large factor is
3C~120.  In the case of 3C 382 a modest enhancement of the continuum
by a factor of a few is possible. For all other objects the jet
inclination angles are such that beaming does not have the desired
effect. Hence the presence a beamed continuum source seems like an
unlikely explanation for the weakness of the lines.

\item{(e)} {\it Low Iron Abundance:}
The observed Fe K$\alpha$ equivalent width decreases with the Fe
abundance, thus a low Fe abundance in the disk could result in weak Fe
lines.  However, there is no good motivation for assuming that the
accretion disk is Fe poor. On the other hand, an Fe deficit as small
as a factor of 2 relative to Seyfert galaxies can bring about the
observed effect. If we take the observed widths of the \feka\ lines in
BLRGs at face value, then the profiles of \feka\ lines of BLRGs are no
different than those of Seyferts.  It could, therefore, be argued that
the entire range of \feka\ $EW$s observed in BLRGs (and Seyferts) can
be attributed to variations in the Fe abundance between objects by a
factor of a few.  If however, the \feka\ lines of BLRGs are
systematically narrower than those of Seyferts, then the Fe abundance
cannot be the only difference between the two clasees of object.

Another aspect of the puzzle of the weak \feka\ line in Pictor~A is
why there is no \feka\ emission from the gas that emits the
double-peaked Balmer lines.  The displaced peaks appear only in the
Balmer lines and {\it not} in the ultraviolet lines (i.e., Ly\a, C\iv,
C\iii]; Eracleous et al., in preparation). In particular the upper
limit to the Ly\a/H\b\ ratio in the displaced peaks is 0.25,
suggesting that the medium where the double-peaked lines originate is
dense with a high column density ($N_{\rm H}\gs$\ten{24-25}~cm\m2) and a low
ionization parameter ($U\ls$\ten{-3}; Halpern et al. 1996;
Collin-Souffrin \& Dumont 1989)\ft{4}. 
\footnote{}{\parindent=0pt \item{\ft{4}} \sml The photoionization models of
Rees, Netzer, \& Ferland (1989), which assume $N_{\rm
H}=10^{23}$~cm\m2 and $U=$\ten{-2} cannot produce Ly\a/H\b\ ratios
less than 10!}  It is therefore paradoxical that the same medium is
not a strong source of
\feka\ emission, unless it subtends a very small solid angle to the
primary X-ray source (e.g., it has the form a thin ring or a pair of
narrow jets).

In the optically thin limit, the observed upper limit to the $EW$ sets the
following limit on the product of the column density of the line-emitting gas
and its covering fraction:
\newcount\QEW \advance\Q by 1 \QEW=\Q
$$
N_{\rm H}\; f_{\rm c} < 1.6\times10^{23} \left({EW\over 140~{\rm eV}}\right)\;
\left({4\times 10^{-5}\over A_{\rm Fe}}\right)~{\rm cm^{-2}},
\eqno{(\the\QEW)}
$$
where $A_{\rm Fe}$ is the iron abundance. Equation (\the\QEW) suggests
that either the covering fraction of the medium responsible for the
double-peaked Balmer lines of order 0.01--0.1, or the Fe abundance is
low. This requirement, along with the abrupt appearance of the
double-peaked Balmer lines, supports the hypothesis that they come
from a region which is dynamically and physically distinct from the
``normal'' broad-line region. Unfortunately, it is difficult to make
more progress in our quest for the origin of the double-peaked lines
before we can distinguish between the two possible reasons for the 
weakness of the \feka\ line mentioned above. Hard X-ray spectra provide
a way out of this impasse, as we discuss briefly in the next section.

\bigskip 
\centerline{\smc 5. conclusions, speculations, and future prospects}
\bigskip

The lack of an \feka\ line in the X-ray spectrum of Pictor~A is
surprising because this line is ubiquitous in the X-ray spectra of
other BLRGs and of Seyfert galaxies. We have considered several
possible explanations for the weakness of the line, none of which is
quite satisfactory, with the possible exception of a low Fe abundance.
An important shortcoming of all other candidate explanations is that
they do not offer a framework in which the \feka\ properties of {\it
all} BLRGs observed with \asca\ can be understood. We have argued that
the extremely broad \feka\ lines observed in some of the BLRGs are a
cause for concern since the velocity of the line wings can exceed the
speed of light. The problem could the consequence of underestimating
the continuum level in the vicinity of the line. This suggestion is
supported by the fact that a re-analysis of the data by Wo\'zniac et
al. (1998) has yielded different results from the original analysis
(i.e., narrower and weaker lines, in general). We have also embarked
on an independent analysis of all of the available data (Sambruna,
Eracleous, \& Mushotzky 1998), in order to verify the findings of
Wo\'zniac et al. (1998).

Assuming that the \feka\ lines of of BLRGs turn out to be weaker than
those of Seyfert galaxies, as the preliminary results of the two
groups, above, suggest, then another one of the candidate explanations
examined in \S4.2 becomes promising: the accretion disks of BLRGs may
have a different structure than those of Seyferts. If reliable
information on the line widths (and profiles) can be obtained from the
\asca\ SIS spectra, it can be used to discriminate between the two
explanations (different disk structure and different Fe abundance). If
the \feka\ lines of BLRGs are systematically narrower than those of
Seyferts, the structure explanation would be favored over the
abundance hypothesis.  Alternatively, hard X-ray spectra can be used
to discriminate between the two possibilities by measuring the
strength of the Compton reflection hump. If the Fe abundance is low,
the optical depth above the Fe K edge at 7.1~keV is small and the
contrast of the Compton reflection hump is higher (e.g., George \& Fabian
1991; Reynolds et al. 1995). If, however, the weakness of the line is
the result of a small covering fraction of the reprocessing medium,
the contrast of the Compton reflection hump decreases along with the
strength of the line. We note in conclusion that a picture in which
the X-ray reprocessing medium in Pictor~A is a narrow ring of gas with
a large column density and a small covering fraction is aesthetically
very pleasing.  Its appeal lies in that it can explain all of the
spectroscopic properties at the same time, including the relative
strengths and profiles of the optical and ultraviolet lines and the
weakness of the \feka\ line.

\bigskip
We are grateful to R. Sambruna and A. Zdziarski for very useful
discussions and the anonymous referee for thoughtful comments.  We
also thank C. Reynolds for sending us the manuscript of his paper on
the \asca\ observations of 3C~111 ahead of
publication. M. E. acknowledges support from Hubble fellowship grant
HF-01068.01-94A from Space Telescope Science Institute, which is
oparated for NASA by the Association of Universities for Research in
Astronomy, Inc., under contract NAS~5-26255. This work was also
supported by NASA grant NAG~5-2524.


\largeskip
\def\ref#1{{\par\noindent \hangindent=3em\hangafter=1 #1\par}}
\def\ditto{\vrule width1.2cm height3pt depth-2.5pt}

\centerline{REFERNCES}
\bigskip

\ref {Alef, W., Preuss, E., Kellerman, K. I., Wu, S. Y., \& Qiu, Y. H. 1994,
      in Compact Extragalactic Radio Sources, NRAO Workshop No. 23,
      eds. J. A. Zensus \& K. I. Kellerman (Green Bank: NRAO), 55}
\ref {Allen, S. W., Fabian, A. C., Idesawa, E., Inoue, H., Kii, T., \& 
      Otani, C. 1997, MNRAS, 286, 765}
\ref {Arnaud, K. A. 1996, in ``Astronomical Data Analysis Software and 
      Systems V'', eds. G. Jacoby \& J. Barnes, ASP Conf. Series, 101, 17}
\ref {Balick, B., Heckman, T. M., \& Crane, P. C. 1982, ApJ, 254, 483}
\ref {Blackburn, J. K., Greene, E. A., \& Pence, B. 1994, User's Guide to 
      FTOOLS (Greenbelt: Goddard Space Flight Center)}     
\ref {Brinkmann, W. \& Siebert, J. 1994, A\&A, 285, 812}
\ref {Chen, K., \& Halpern, J. P. 1989, ApJ, 344, 115}
\ref {Collin-Souffrin, S., \& Dumont, A. M. 1989, A\&A, 213, 39}
\ref {Corbin, M. R. 1997, ApJ, 485, 517}
\ref {Eracleous, M. \& Halpern, J. P. 1994, ApJS, 90, 1}
\ref {\ditto\ . 1998$a$, in preparation} 
\ref {\ditto\ . 1998$b$, in ``Accretion Processes in Astrophysical Systems:
      Some Like it Hot'', eds. S. S. Holt \& T. R. Kallman (New York: ASP), in
      press} 
\ref {Eracleous, M., Halpern, J. P., \& Livio, M. 1996, ApJ, 459, 89}
\ref {Fabian, A. C., Rees, M. J., Stella, L., \& White, N. E. 1989, MNRAS, 
      238, 729}
\ref {George, I. M., \& Fabian, A. C., 1991, MNRAS, 249, 352}
\ref {George, I. M., Nandra, K., \& Fabian, A. C. 1990, MNRAS, 242, 28P}
\ref {Grandi, P., Sambruna, R. M., Maraschi, L., Matt, G., Urry, C. M., 
      \& Mushotzky, R. F., 1997, ApJ, 487, 636}
\ref {Guilbert, P. W., \& Rees, M. J. 1988, MNRAS, 233, 475}
\ref {Haardt, F., \& Maraschi, L. 1991, ApJ, 308, L51}
\ref {\ditto\ . 1993, ApJ, 413, 507}
\ref {Haardt, F., Maraschi, L., \& Ghisellini, G. 1994, ApJ, 432, L95}
\ref {Halpern, J. P., Eracleous, M., Filippenko, A. V., \& Chen, K.
      1996, ApJ, 467, 704}
\ref {Halpern, J. P., \& Eracleous, M. 1994, ApJ, 433, L17}
\ref {Heiles, C. \& Cleary, M. N. 1979, AuJPA, 47, 1}
\ref {Ingham, J. 1994, The XSELECT User's Guide
      (Greenbelt: Goddard Space Flight Center)}     
\ref {Jackson, N., \& Browne, I. W. A. 1991, MNRAS, 250, 414}
\ref {Jackson, N., Penston, M. V., \& P\'erez, E. 1991, MNRAS, 248, 577}
\ref {Jones, P. A., \& McAdam, W. B. 1992, ApJS, 80, 137}
\ref {Kruper, J. S., Urry, C. M., \& Canizares, C. R. 1990, ApJS, 74, 347}
\ref {Lightman, A. P., \& White, T. R. 1988, ApJ, 335, 57}
\ref {Magdziarz, P., \& Zdziarski, A. A. 1995, MNRAS, 273, 837} 
\ref {Matt, G., Perola, G. C., \& Piro, L. 1991, A\&A, 247, 2}
\ref {McCarthy, P., van~Breugel, W., \& Kapahi, V. J. 1991, ApJ, 371, 478}
\ref {Miley, G. K., \& Miller, J. S. 1979, ApJ, 228, L55}
\ref {Morrison, R. \& McCammon, D. 1983, ApJ, 270, 119}
\ref {Mushotzky, R. F. et al. 1995, MNRAS, 272, P9}
\ref {Nandra, K., \& George, I. M. 1994, MNRAS, 267, 974}
\ref {Nandra, K., George, I. M., Mushotzky, R. F., Turner, T. J., \& 
      Yaqoob, T. 1997$a$, ApJ, 477, 602}
\ref {\ditto\ . 1997$b$, ApJ, 488, L91}
\ref {Nandra, K., \& Pounds, K. A. 1994, MNRAS, 268, 405}
\ref {Narayan, R. \& Yi, I. 1994, ApJ, 428, L13}
\ref {Narayan, R. \& Yi, I. 1995, ApJ, 444, 231}
\ref {Nillson, K., Valtonen, M. J., Kotilainen, J., \& Jaakkola, T. 1993,
      ApJ, 413, 453}
\ref {Pounds, K. A., Nandra, K., Stewart, G. C., \& Leighly, K. 1989, MNRAS,
      240, 769}
\ref {Rees, M. J., Begelman, M. C., Blandford, R. D., \& Phinney, E. S. 1982, 
      Nature, 295, 17}
\ref {Reynolds, C. S. 1997, MNRAS, 286, 513}
\ref {Reynolds, C. S., Iwasawa, K., Crawford, C. S., \& Fabian, A. C. 
      1997, MNRAS, in press}
\ref {Reynolds, C. S., Fabian, A. C., \& Inoue, H. 1995, MNRAS, 276, 1311}
\ref {Sambruna, R. M., Eracleous, M., \& Mushotzky, R. F. 1998, in preparation}
\ref {Sambruna, R. M., George, I. M., Mushotzky, R. F., Nandra, K., \& 
      Turner, T. J. 1997, ApJ, in press}
\ref {Sincell, M. W. \& Krolik, J. H. 1997, ApJ, 476, 605}
\ref {Singh, K. P., Rao, A. R., \& Vahia, M. N. 1990, MNRAS, 246, 706}
\ref {Steiner, J. E. 1981, ApJ, 250, 469}
\ref {Sulentic, J. W., Marziani, P., Zwitter, T., \& Calvani, M. 1995, ApJ,
      438, L1}
\ref {Tanaka, Y., Inoue, H., \& Holt, S. S. 1994, PASJ, 46, L37}
\ref {Tanaka, Y., et al. 1995, Nature, 375, 659}
\ref {Turner, T. J. \& Pounds, K. A. 1989, MNRAS, 232, 463}
\ref {Vermuelen, R. C., \& Cohen, M. H. 1994, ApJ, 430, 467}
\ref {Walker, R. C., Walker, M. A., \& Benson, J. M. 1988, ApJ, 335, 668}
\ref {Wehrle, A. E., Cohen, M. H., Unwin, S. C., Aller, H. D., Aller, M. F., 
      \& Nicolson, G. 1992, ApJ, 391, 589}
\ref {Wills, B. J., \& Browne, I. W. A. 1986, ApJ, 302, 56}
\ref {Wo\'zniac, P. R., Zdziarski, A. A., Smith, D., Madejski, G. M., \& Johnson,
      W. N. 1998, MNRAS, in press}
\ref {Yaqoob, T. 1996, in ``Minutes of \asca\ Calibration Meeting, ISAS, 
      1996 Nov. 16''}
\ref {Yaqoob, T., McKernan, B., Done, C., Serlemitsos, P. J., \&
      Weaver, K. A. 1993, ApJ, 416, L5}
\ref {Zdziarski, A. A., Johnson, W. N., Done, C., Smith, D., \&
      McNaron-Brown, K. 1995, ApJ, 438, L63}
\ref {Zensus, J. A. 1989, in Lecture Notes in Physics vol. 334 ``BL Lac
      Objects''. eds. L. Maraschi, T. Maccacaro, \& M.-H. Ulrich (Berlin:
      Springer), 3}


\end